\documentclass[prc,preprint,showpacs,preprintnumbers,amsmath,amssymb]{revtex4}

\usepackage[mathscr]{eucal}
\usepackage{graphicx}
\def\slr#1{\setbox0=\hbox{$#1$}           
   \dimen0=\wd0                                 
   \setbox1=\hbox{/} \dimen1=\wd1               
   \ifdim\dimen0>\dimen1                        
      \rlap{\hbox to \dimen0{\hfil/\hfil}}      
      #1                                        
   \else                                        
      \rlap{\hbox to \dimen1{\hfil$#1$\hfil}}   
      /                                         
   \fi}

\def\be{\begin{eqnarray}}
\def\ee{\end{eqnarray}}

\renewcommand{\theequation}%
    {\arabic{section}.\arabic{equation}}
\makeatletter \@addtoreset{equation}{section} \makeatother

\begin{document}

\preprint{BCCNT: 04/111/329}

\title{The Gluon Propagator in Minkowski and Euclidean Space: \\Role of an $A^2$ Condensate}

\author{Xiangdong Li}
\affiliation{%
Department of Computer System Technology\\
New York City College of Technology of the City University of New
York\\
Brooklyn, New York 11201 }%

\author{C. M. Shakin}
\email[email address:]{casbc@cunyvm.cuny.edu}

\affiliation{%
Department of Physics and Center for Nuclear Theory\\
Brooklyn College of the City University of New York\\
Brooklyn, New York 11210
}%

\date{November, 2004}

\begin{abstract}
In a recent work we described the form taken by the
Minkowski-space gluon propagator in the presence of an $<A^a_\mu
A^\mu_a>$ condensate. This condensate has been seen to play an
important role in calculations which make use of the operator
product expansion. The latter work has led to the publication of a
large number of papers which discuss how the $<A^a_\mu A^\mu_a>$
condensate could play a role in a gauge-invariant formulation. In
the present work we discuss the properties of both the
Euclidean-space and Minkowski-space gluon propagator. In the case
of the Euclidean-space propagator we can make contact with the
results of QCD lattice calculations of the propagator in the
Landau gauge. With an appropriate choice of normalization
constants, we present a unified representation of the gluon
propagator that describes both the Minkowski-space and
Euclidean-space dynamics in which the $<A_\mu^aA_a^\mu>$
condensate plays an important role.

\end{abstract}


\maketitle


\section{introduction}

Recently, studies making use of the operator product expansion
(OPE) have provided evidence for the importance of the condensate
$<A_\mu^aA_a^\mu>$ [1-3]. (There is a suggestion that such a
condensate may be related to the presence of instantons in the
vacuum [4].) The importance of that condensate raises the question
of gauge invariance and there are now a large number of papers
that address that and related issues [5-19]. We will not attempt
to review that large body of literature, but will consider how the
presence of an $<A_\mu^aA_a^\mu>$ condensate affects the form of
the gluon propagator. We may mention the work of Kondo [7] who was
responsible for introducing a BRST-invariant condensate of
dimension two, \be \mathcal{Q}=\frac{1}{\Omega}<\int
d^4x\,\mbox{Tr}\left(\frac{1}{2}A_\mu(x)A_\mu(x)-\alpha ic(x)\cdot
\bar{c}(x)\right)>, \ee where $c(x)$ and $\bar{c}(x)$ are
Faddeev-Popov ghosts, $\alpha$ is the gauge-fixing parameter and
$\Omega$ is the integration volume. Kondo points out that $\Omega$
reduces to $A_{min}^2$ in the Landau gauge, $\alpha=0$. The
minimum value of the integrated squared potential is $A^2_{min}$,
which has a definite physical meaning [7].

In a recent work we considered the Minkowski-space gluon
propagator in the presence of an $A^2$ condensate [20]. We found
that the propagator has no on-mass-shell poles, so that the gluon
was a nonpropagating mode in the presence of the vacuum condensate
[21]. The form we obtained for the propagator was \be
D^{\mu\nu}(k)=\left(g^{\mu\nu}-\frac{k^\mu k^\nu}{k^2}\right)D(k)
\ee with \be
D(k)=\frac{Z_1}{k^2-m^2+\frac{4}{3}\frac{k^2m^2}{k^2-m^2}}.\ee
Here $Z_1$ is a normalization parameter which we put equal to 3.82
so that we may obtain a continuous representation as we pass from
Minkowski to Euclidean space.  In Fig. 1 we show $D(k)$ with
$m^2=0.25$ GeV$^2$. (We remark that $D(k)=0$ when $k^2=m^2$,
$D(k)=-Z_1/m^2$ at $k^2=0$, and $D(k)\rightarrow Z_1/k^2$ for
large $k^2$.) If we choce $Z_1=15.28m^2=3.82$ our result for the
propagator will be continuous at $k^2=0$ when we consider both the
Euclidean-space and Minkowski-space propagators.

The organization of our work is as follows. In Section II we
discuss the Euclidean-space gluon propagator as obtained in a
lattice simulation of QCD in the Landau gauge and in the absence
of quark degrees of freedom [22]. (We also record in the Appendix
a number of semi-phenomenological analytic forms which are meant
to represent the Euclidean-space propagator.) In Section III we
summarize the results of our analysis and provide some additional
discussion.

\section{QCD lattice calculations and phenomenological forms for the Euclidean-space gluon propagator}

Results for the gluon propagator obtained in a lattice simulation
of QCD are given in Ref. [22]. In that work the authors also
record several phenomenological forms. We reproduce these forms in
the Appendix for ease of reference. Of these various forms we will
make use of model A of Ref. [22] which has the form \be
D^L(k^2)=Z\left[\frac{AM^{2\alpha}}{(k^2+M^2)^{1+\alpha}}+\frac{1}{k^2+M^2}L(k^2,M)\right],\ee
with \be L(k^2,M)\equiv
\left[\frac{1}{2}\ln(k^2+M^2)(k^{-2}+M^{-2})\right]^{-d_D},\ee and
$d_D=13/22$. The parameters used in Ref. [22] to provide a very
good fit to the QCD lattice data are \be Z=2.01^{+4}_{-5},\ee \be
A=9.84^{+10}_{-86},\ee \be M=0.54^{+5}_{-5},\ee and \be
\alpha=2.17^{+4}_{-19}.\ee Note that $M$ in GeV units is $1.018$
GeV. Rather than work with the lattice data we will use Eqs.
(2.1)-(2.6) when we compare our results with the lattice data. In
Fig. 2 we show $k^2D^L(k)$ of Eq. (2.1) and in Fig. 3 we show
$D^L(k)$. These functions are represented by the solid lines in
Figs. 2 and 3. Note that Eq. (1.3) may be written in Euclidean
space as \be
D_E(k)=-\frac{Z_1}{k^2_E+m^2-\frac{4}{3}\frac{k^2_Em^2}{k^2_E+m^2}}.\ee
This form is useful for $k_E^2<1$ GeV$^2$ and we therefore
consider various phenomenological forms which may be used to
extend Eq. (2.7) so that we may attempt to fit the lattice result
over a broader momentum range. To that end, we make use of Ref.
[23]. The authors of that work define the Landau gauge gluon
propagator as \be
<A^a_\mu(k)A^a_\nu(k')>=V\delta(k+k')\delta^{ab}\left(\delta_{\mu\nu}-\frac{k_\mu
k_\nu}{k^2}\right)\frac{Z(k^2)}{k^2}, \ee with \be
Z(k^2)=\omega\left(\frac{k^2}{\Lambda^2_{QCD}+k^2}\right)^{2\kappa}(\alpha(k^2))^{-\gamma},\ee
and $\gamma=-13/22$. (We do not ascribe any particular
significance to Eq. (2.9). We use Eq. (2.9) as a phenomenological
form which could be replaced by a form which provides a better fit
to the data within the context of our model at some future time.
We believe Eq. (2.9) is useful, since it is a simple matter to
remove the first term of that equation and introduce a propagator
that has the small $k^2$ behavior of our model.)

The authors of Ref. [23] introduce two choices for $\alpha(k^2)$
of Eq. (2.9). We use their form for $\alpha_2(k^2)$: \be
\alpha_2(k^2)=\frac{\alpha(0)}{\ln\left[e+a_1\left(\frac{k^2}{\Lambda^2_{QCD}}\right)^{a_2}\right]}.\ee
In their analysis they put $\kappa=0.5314$, $\Lambda_{QCD}=354$
MeV, $\alpha(0)=2.74$, $a_1=0.0065$ and $a_2=2.40$. (Here, we have
not recorded the uncertainties in these values which are given in
Table 2 of Ref. [23].) As we proceed, we will change these values
somewhat. As a first step we remove the first factor in Eq. (2.9)
and write \be Z(k^2)=Z_2(\alpha_2(k^2))^{-\gamma}.\ee We now use
$a_1=0.0080$ and $a_2=2.10$ rather than the values given above. In
Fig. 4 we show $(\alpha_2(k))^{13/22}$ as a function of $k$, using
our modified values of $a_1$ and $a_2$.

We now define \be
D_E(k_E)=-\frac{Z_2(\alpha(k^2))^{-\gamma}}{k_E^2+m^2-\frac{4}{3}\frac{k_E^2m^2}{k_E^2+m^2}}.\ee
The function $-k^2D_E(k_E)$ is shown in Fig. 2 as a dotted line.
In this calculation we have put $Z_2=2.11$. We find a good
representation of the lattice result for $k_E<2$ GeV,

In Fig. 3 we compare $D_E(k_E)$ with the result of the lattice
calculation which is represented by the solid line. In Fig. 5 we
combine our results in Minkowski and Euclidean space and show the
values of $k^2D(k^2)$ for both positive and negative $k^2$ values.
For positive $k^2$ we use $D(k)$ of Eq. (1.3) and for negative
values of $k^2$ we use $D_E(k_E^2)$ of Eq. (2.12). Equality of
these functions at $k^2=0$ implies $Z_1=Z_2(\alpha(0))^{13/22}$,
or $Z_1=1.81Z_2$. (In our work we have used $Z_1=3.82$ and
$Z_2=2.11$. See Eqs. (1.3) and (2.12).) In Fig. 6 we show $D(k^2)$
rather than $k^2D(k^2)$, which was shown in Fig. 5.

\begin{figure}
\includegraphics[bb=40 25 200 200, angle=0, scale=1]{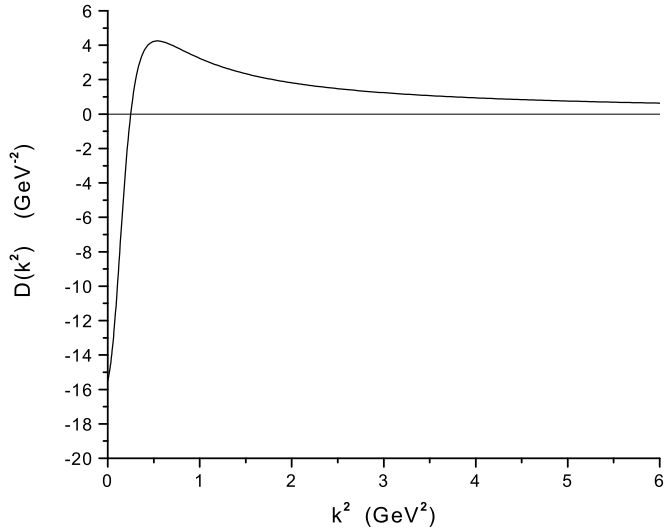}%
\caption{The function $D(k^2)$ of Eq. (1.3) is shown in Minkowski
space. The value for large $k^2$ is given by $Z_1/k^2$ with
$Z_1=3.87$. Here $m=0.50$ GeV.}
\end{figure}

\begin{figure}
\includegraphics[bb=40 25 240 240, angle=0, scale=1]{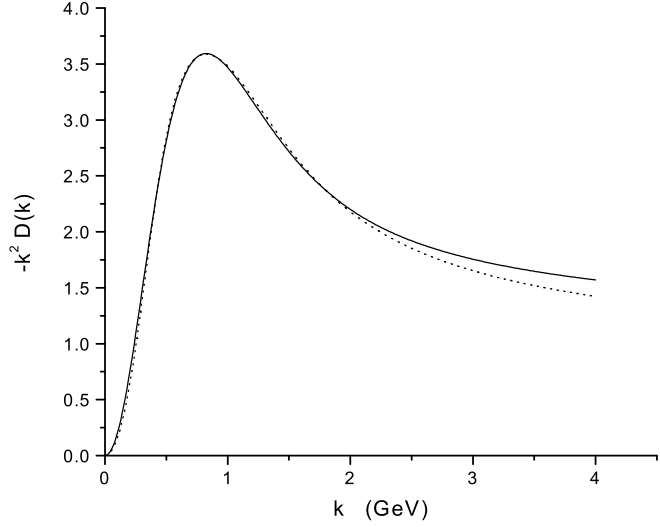}%
\caption{The function $-k_E^2D_E(k)$ is shown. The solid line
represents the QCD lattice data, while the dotted line represents
$-k^2_ED_E(k)$ in the case that $D_E(k)$ is given in Eq. (2.12).}
\end{figure}

\begin{figure}
\includegraphics[bb=40 25 240 240, angle=0, scale=1]{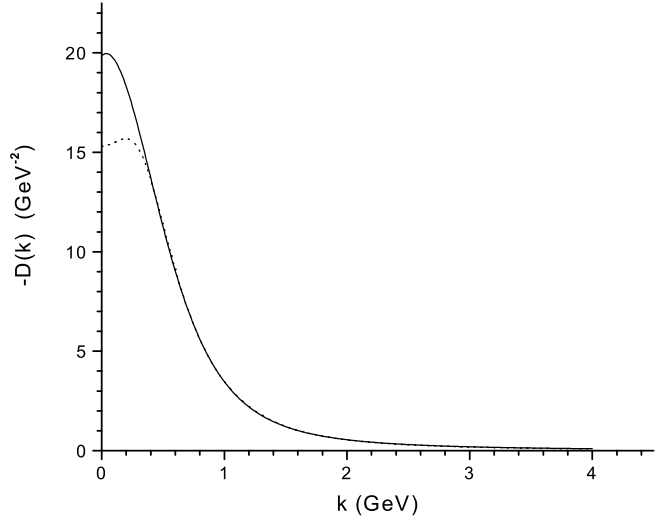}%
\caption{The function $-D_E(k)$ is shown. The solid line
represents the QCD lattice data, while the dotted line represents
$-D_E(k)$ of Eq. (2.12). [See Fig. 2.]}
\end{figure}

\begin{figure}
\includegraphics[bb=40 25 240 240, angle=0, scale=1]{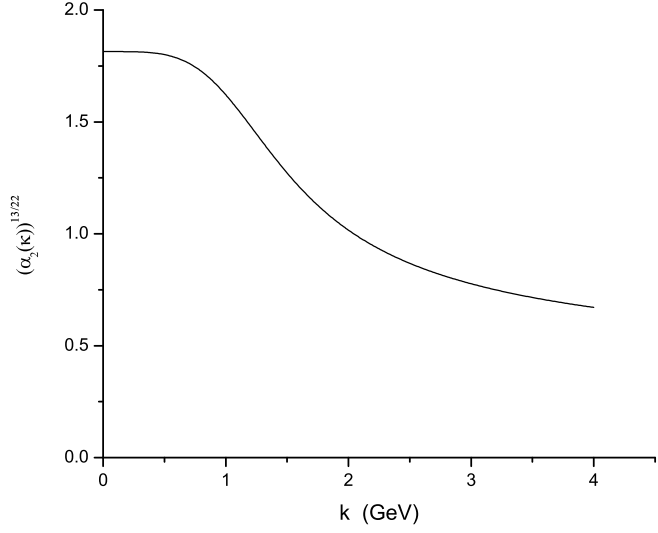}%
\caption{The function $(\alpha_2(k))^{13/22}$ is shown. [See Eq.
(2.10).] Note that $(\alpha_2(0))^{13/22}=1.81$.}
\end{figure}

\begin{figure}
\includegraphics[bb=40 25 240 240, angle=0, scale=1]{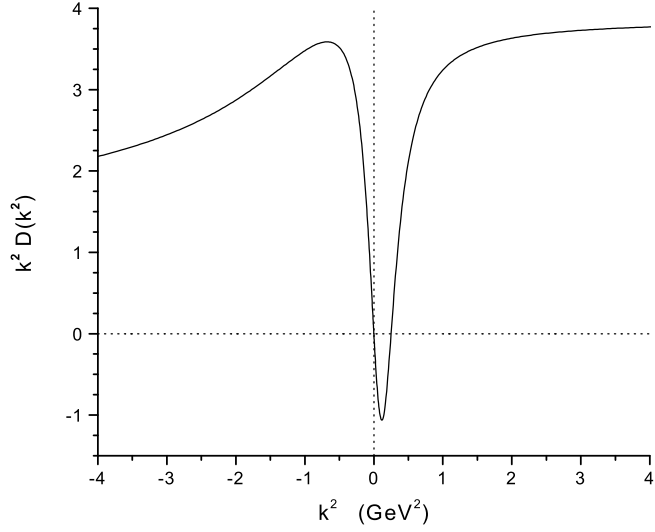}%
\caption{For $k^2>0$ the solid line represents $k^2D(k^2)$ with
$D(k^2)$ given by Eq. (1.3). Here, $Z_1=3.82$. For $k^2<0$ we show
$k^2D_E(k^2)$, where $D_E(k^2_E)$ is given by Eq. (2.12) with
$Z_2=2.11$.}
\end{figure}

\begin{figure}
\includegraphics[bb=40 25 240 240, angle=0, scale=1]{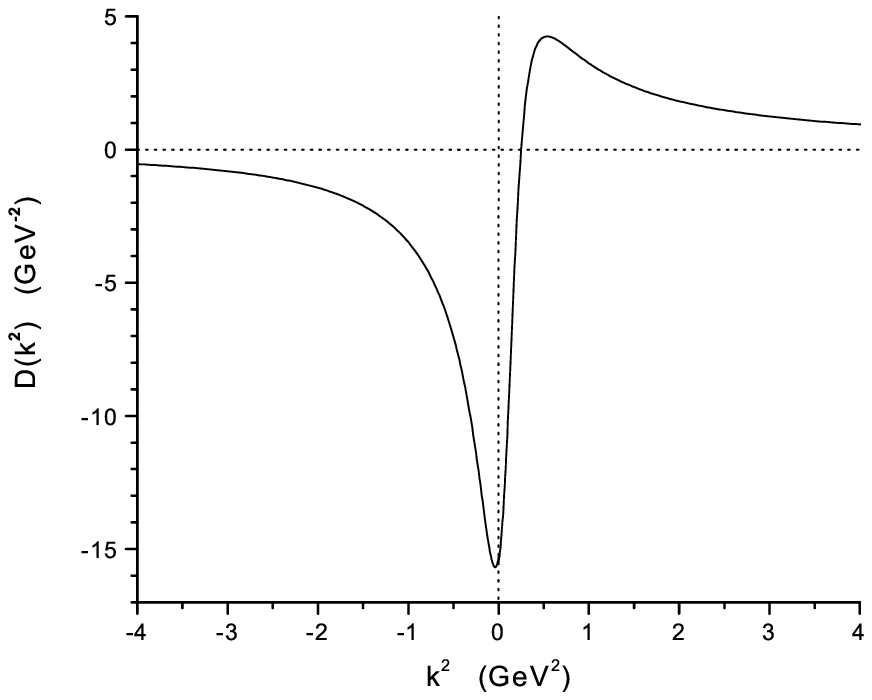}%
\caption{Same as Fig. 5 except that $D(k^2)$ is shown.}
\end{figure}

  \section{Discussion and Conclusions}

In this work we have provided a representation of the gluon
propagator in both Euclidean and Minkowski space. The
Minkowski-space propagator has only complex poles and that implies
that the gluon is a nonpropagating mode in the QCD vacuum. Our
analysis takes into account the important condensate $<A_a^\mu
A_\mu^a>$ which is responsible for mass generation for the gluon.
Our work has some relation to that of Cornwall [24] who obtained a
gluon mass of $500\pm 200$ MeV in his analysis. Cornwall also
suggested that ``quark confinement arises from a vertex condensate
supported by a mass gap."

In recent work Gracey obtained a pole mass of the gluon of
$2.13\Lambda_{\overline{MS}}$ in a two-loop renormalization scheme
[25]. If we put $\Lambda_{\overline{MS}}=250$ MeV, the mass
obtained at two-loop order in Ref. [25] is $532$ MeV, which is
close to the value of $500$ MeV used in the present work. (We
remark that in Ref. [21] we obtained a gluon mass of $530$ MeV, if
we made use of Eq. (3.18) of that reference, which includes the
effect of including various exchange terms in our analysis of the
relevant matrix elements.)

\appendix
  \renewcommand{\theequation}{A\arabic{equation}}
  \setcounter{equation}{0}  
  \section{}  

For ease of reference we record various semi-phenomenological
forms which are meant to represent the Euclidean-space gluon
propagator.

Gribov [26]: \be D^L(k^2)=\frac{Zk^2}{k^4+M^4}L(k^2,M).\ee Stingl
[27]: \be D^L(k^2)=\frac{Zk^2}{k^4+2A^2k^2+M^4}L(k^2,M).\ee
Marenzoni et al. [28]: \be
D^L(k^2)=\frac{Z}{(k^2)^{1+\alpha}+M^2}.\ee Cornwall I [24]: \be
D^L(k^2)=Z\left[[k^2+M^2(k^2)]\ln\left(\frac{k^2+4M^2(k^2)}{\Lambda^2}\right)\right]^{-1},
\ee where \be M(k^2)=M
\left[\frac{\ln\left(\frac{k^2+4M^2}{\Lambda^2}\right)}{\ln\left(\frac{4M^2}{\Lambda^2}\right)}\right]^{-6/11}.\ee
Cornwall II [29]: \be
D^L(k^2)=Z\left[[k^2+M^2]\ln\left(\frac{k^2+4M^2}{\Lambda^2}\right)\right]^{-1}.
\ee Cornwall III [29]: \be
D^L(k^2)=\frac{Z}{k^2+Ak^2\ln\left(\frac{k^2}{M^2}\right)+M^2}.\ee
Model A [22]: \be D^L(k^2)=
Z\left[\frac{AM^{2\alpha}}{(k^2+M^2)^{1+\alpha}}+\frac{1}{k^2+M^2}L(k^2,M)\right].\ee
The parameters for model A are given in Eqs. (2.3)-(2.6).\\Model B
[22]: \be D^L(k^2)=
Z\left[\frac{AM^{2\alpha}}{(k^2)^{1+\alpha}+(M^2)^{1+\alpha}}+\frac{1}{k^2+M^2}L(k^2,M)\right].\ee
Model C [22]: \be D^L(k^2)=
Z\left[\frac{A}{M^2}e^{-(k^2/M^2)^\alpha}+\frac{1}{k^2+M^2}L(k^2,M)\right].\ee


\end{document}